\documentclass[mathleft]{an}
\usepackage{graphicx}
\usepackage{times}
\usepackage{natbib}
\bibpunct{(}{)}{;}{a}{}{,}
\def\aap{A\&A}
\def\nat{Nature}
\def\nar{NewAR}
\begin{document}

% The following seven commands are intended for editorial usage and should be ignored by
% the author(s).
\Pagespan{6}{}
\Yearpublication{}
\Yearsubmission{}
\Month{}  
\Volume{}
\Issue{}
\DOI{doi}

\title{An Intermediate-Mass Black Hole candidate in M51?}

\author{H.M. Earnshaw\inst{1}\fnmsep\thanks{\email{hannah.earnshaw@durham.ac.uk}}
}
\titlerunning{An IMBH candidate in M51?}
\authorrunning{H.M. Earnshaw}
\institute{Centre for Extragalactic Astronomy, Durham University, South Road, Durham, DH1 3LE, UK}

\received{}
\accepted{}
\publonline{}


\abstract{We present the current results of an investigation into M51 ULX-7, using archival data from {\it XMM-Newton}, {\it Chandra} and {\it NuSTAR}, and optical and radio data from {\it HST} and {\it VLA}. The source has a consistently hard power-law X-ray spectrum and high short-term variability. This is unusual variability behaviour for a ULX, as we would expect highly variable ULXs to have soft energy spectra. The power spectrum features a break at $\sim10^{-3}$\,Hz, from low frequency spectral index $\alpha=0.1$ to high frequency spectral index $\alpha=0.8$, analogous to the low frequency break found in power spectra of black holes accreting in the low/hard state. We do not observe a corresponding high frequency break, however taking the white noise level as a frequency lower  limit of the break, we can calculate a black hole mass upper limit of $9.12\times10^{4}$\,M$_{\odot}$, assuming that the ULX is in the low/hard state. While there is no radio detection, we find a flux density upper limit of 87 $\mu$Jy/beam. Using the X-ray/radio fundamental plane, we calculate a black hole mass upper limit of $1.95\times10^{5}$\,M$_{\odot}$. Therefore, this ULX is consistent with being an IMBH accreting in the low/hard state.}

\maketitle

\section{Introduction}

Ultra-luminous X-ray sources (ULXs) are non-nuclear point sources with X-ray luminosity $L_{\rm X}>10^{39}$\,erg\,s$^{-1}$, the Eddington luminosity for black holes (BHs) of typical mass $\sim10$\,M$_{\odot}$ \citep{roberts07, feng11}. Such a luminosity can only be reached by a number of extreme scenarios. The BH could be very massive, with $10^2~<~M_{\rm BH}~<~10^5$\,M$_{\odot}$, known as an intermediate-mass BH (IMBH). This provides a simple physical explanation for the high luminosity, but it is challenging to explain the formation of such objects in the quantities that we observe using this model. Alternatively, such high luminosities could be reached through extreme accretion scenarios such as super-Eddington accretion and/or geometric beaming.

Recent studies of high-quality ULX spectra have shown that they exhibit different spectral features to lower-luminosity BHs, featuring either a broadened disc or a two-component spectrum with a soft excess and a turnover at energies above 3--5\,keV (e.g. \citealt{stobbart06, gladstone09}). This latter spectral shape could represent a super-Eddington accretion state called the hard or soft ultraluminous state, depending on which component dominates the emission \citep{sutton13}. Of these two ultraluminous states, high amounts of variability are only seen in the hard component of the soft ultraluminous state \citep{middleton15}. The hard ultraluminous state tends to show very little variability, in constrast with the high amounts of variability seen in the sub-Eddington low/hard state. Both ultraluminous states can be explained by a model incorporating a very hot inner disc and a cooler, clumpy outflowing wind. Which component is dominant depends on the inclination, and the clumpy wind imprints variability onto the hard emission at low inclinations as it obscures the inner source.

While the many ULXs can be explained in this way, a small number of sources are sufficiently luminous ($L_{\rm X}>10^{41}$\,erg\,s$^{-1}$) that even super-Eddington accretion cannot fully explain their extreme luminosities. Such sources, such as ESO 243-49 HLX-1 \citep{farrell09}, are currently the best candidates for potential IMBHs.

\subsection{A Catalogue of ULXs and M51 ULX-7}

We have created a new, clean catalogue of ULXs (\citeauthor{earnshaw16} in prep.b) using data from 3XMM-DR4 \citep{xmm} matched with the Third Reference Catalog of Bright Galaxies \citep{rc3}. This catalogue contains 331 ULXs, many of which do not appear in previous catalogues. We used this sample as a starting point to locate sources of interest such as variable ULXs, as they are relatively uncommon. Within the catalogue are 10 sources that were flagged as variable, including M51 ULX-7 (henceforth ULX-7) which is of especial interest due to its hard spectrum, putting it at odds with the low-variability hard ultraluminous state.

Previous studies have found ULX-7 to be variable, and while it was initially thought to possibly be a periodic source \citep{liu02}, later research indicates that its variability is aperiodic, resulting from stochastic processes (e.g. \citealt{terashima06}). As of yet no in-depth investigation into this source has been undertaken, however a good deal of archival data from {\it XMM-Newton} and {\it Chandra} already exists, as well as an observation with {\it NuSTAR}, and optical and radio data. Therefore it is an ideal source to study in order to discover more about the extremes of BH accretion (\citeauthor{earnshaw15} in prep.a).

We present the results of our investigation at the time of the 2015 {\it XMM-Newton} science workshop, and discuss the possible interpretations of this unusual ULX.

\section{Results From Archival Observations}

As a face-on spiral galaxy with a high amount of star formation and X-ray activity, M51 has been the subject of a large number of observations across multiple wavelengths, therefore there is a good amount of archival data which we can use to examine ULX-7 from a multi-wavelength perspective. The source position is 13:30:01.0~+47:13:44 \citep{kilgard05}.

\subsection{Radio}

The presence of radio data is highly dependent on the state of the BH being observed. Radio jets would be expected from a source in the low/hard state. We used archival 1.5\,GHz {\it VLA} data from project 11A142, August 2011, to search for a detection at the location of the ULX. No detection was found, so we took the rms at $8.73\times10^{-5}$\,Jy\,beam$^{-1}$ to calculate an upper limit to the radio flux density of 87\,$\mu$Jy\,beam$^{-1}$.

\subsection{Optical}

\begin{figure}
\centering
\includegraphics[width=70mm]{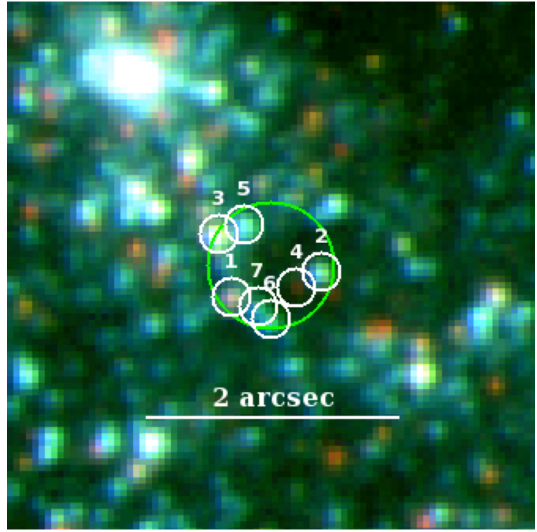}
\caption{True-colour optical image of the region surrounding ULX-7 using {\it HST} data, with the F814W band in red, F555W in green and F435W in blue. The source position is marked by a 0.5\,arcsec radius error circle, and the seven marked couterparts correspond to the counterparts plotted in Fig.~\ref{fig:colmag}.}
\label{fig:opt}
\end{figure}

M51 was observed with the {\it HST} ACS/WFC instrument in 2005, as part of the Hubble Heritage project. We used the pre-processed images from the Hubble Legacy Archive to search for possible optical counterparts for ULX-7 within 0.5\,arcsec of the source position. Using the PSF-fitting routine DAOPHOT II/ALLSTAR \citep{daophot}, we are able to characterise seven potential counterparts. Visual inspection shows there to be further potential counterparts, but they are too faint to characterise. The counterparts are marked in Fig.~\ref{fig:opt}, and are located near to a young star cluster seen in the top left.

\begin{figure}
\centering
\includegraphics[width=80mm]{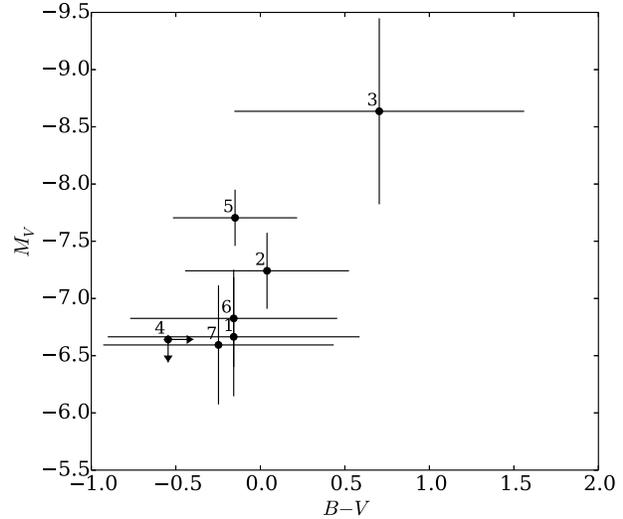}
\caption{The {B-V} colour plotted against the $V$-band absolute magnitude for the seven optical counterparts marked in Fig.~\ref{fig:opt}. Errors are calculated based on the estimated standard error from the DAOPHOT results.}
\label{fig:colmag}
\end{figure}

We show the $B-V$ colour plotted against the $V$-band absolute magnitude for the potential counterparts in Fig.~\ref{fig:colmag}. Most of the counterparts have colours consistent with being OB supergiants (possibly including some disc irradiation), as expected from the young stellar environment, and would thus be suitable companions for a high-mass X-ray binary (HMXB) system. The exception is counterpart 3, which is much brighter and redder than the other objects - this makes it likely to be a cluster of lower mass stars.

The colours and magnitudes alone cannot rule out a background AGN, however we also measured the ratio between the maximum X-ray flux and the $V$-band flux. For all potential counterparts, we found $F_{\rm X}/F_{\rm opt}>10$. Since the majority of AGN have an X-ray/optical flux ratio of $F_{\rm X}/F_{\rm opt}<10$, it is unlikely that any of these counterparts are a background source. This applies even to the potential counterparts we are unable to characterise, as they will have lower optical fluxes and thus higher X-ray/optical ratios.

\subsection{X-ray spectra}

We are able to make use of archival observations from {\it XMM-Newton}, {\it Chandra} and {\it NuSTAR} to characterise the spectrum of ULX-7. M51 has been observed six times with {\it XMM-Newton} between 2003 and 2011, twelve times with {\it Chandra} between 2000 and 2012, and once with {\it NuSTAR} in 2012.

Inspection of the {\it XMM-Newton} images of the source show that it is found within extended soft X-ray emission in the host galaxy, which we need to account for correctly to determine the shape of the intrinsic source spectrum. Therefore we first examine the {\it Chandra} spectrum of the gas within an annulus with inner radius of 3\,arcsec and outer radius of 20\,arcsec surrounding the source. We fit the diffuse emission spectrum simultaneously across the five longest {\it Chandra} observations. We find that the emission is well-fitted by two {\tt mekal} components, which is consistent with previous studies of the diffuse emission in M51 (e.g. \citealt{owen09}). The temperatures of the two gas components are given in Table~\ref{tab:mekal}. 

\begin{table}
\centering
\caption{The {\tt mekal} component temperatures and goodness of fit for the extended diffuse emission around ULX-7.}
\label{tab:mekal}
\begin{tabular}{ccc}\hline
kT$_1$ & kT$_2$ & $\chi^2$/d.o.f. \\ 
(keV) & (keV) & \\
\hline
$0.26\pm0.04$ & $0.8\pm0.2$ & 279.9/258 \\
\hline
\end{tabular}
\end{table}

Next, we fitted the {\it Chandra} and {\it XMM-Newton} source spectra, extracted from 3\,arcsec and 20\,arcsec circular regions around the source respectively, with an absorbed power-law model ({\tt tbabs*powerlaw} in {\sc xspec}). In the case of the {\it XMM-Newton} spectra we also added the two-component gas spectrum with fixed temperatures and normalisations to the fit ({\tt mekal+mekal+tbabs*powerlaw}). By accounting for soft contamination by the diffuse emission this way, we found that most of the {\it XMM-Newton} and {\it Chandra} spectra were well-fitted by a simple absorbed power-law model with $\Gamma\sim1.5$, regardless of flux. In some cases, the fit  was not excellent, however the residuals did not indicate an unfitted component, but rather a noisy spectrum due to a low amount of data. The one exception was the longest {\it XMM-Newton} observation 0303420101, for which there was a marginal improvement of fit of $\Delta\chi^2\sim17$ for 2 fewer degrees of freedom with the addition of a multicolour disc model ({\tt diskBB}). 

This is all in contrast to the expected spectrum of a source in the ultraluminous state, which features a soft excess and a turnover at energies greater than 3--5\,keV. However, these features are only seen in very high quality spectra \citep{gladstone09}, so we may not expect to see them anyway, regardless of the underlying state. However, the presence of {\it NuSTAR} data allows us to probe higher energies and verify the existence of a turnover if one is present. 

\begin{figure}
\centering
\includegraphics[width=80mm]{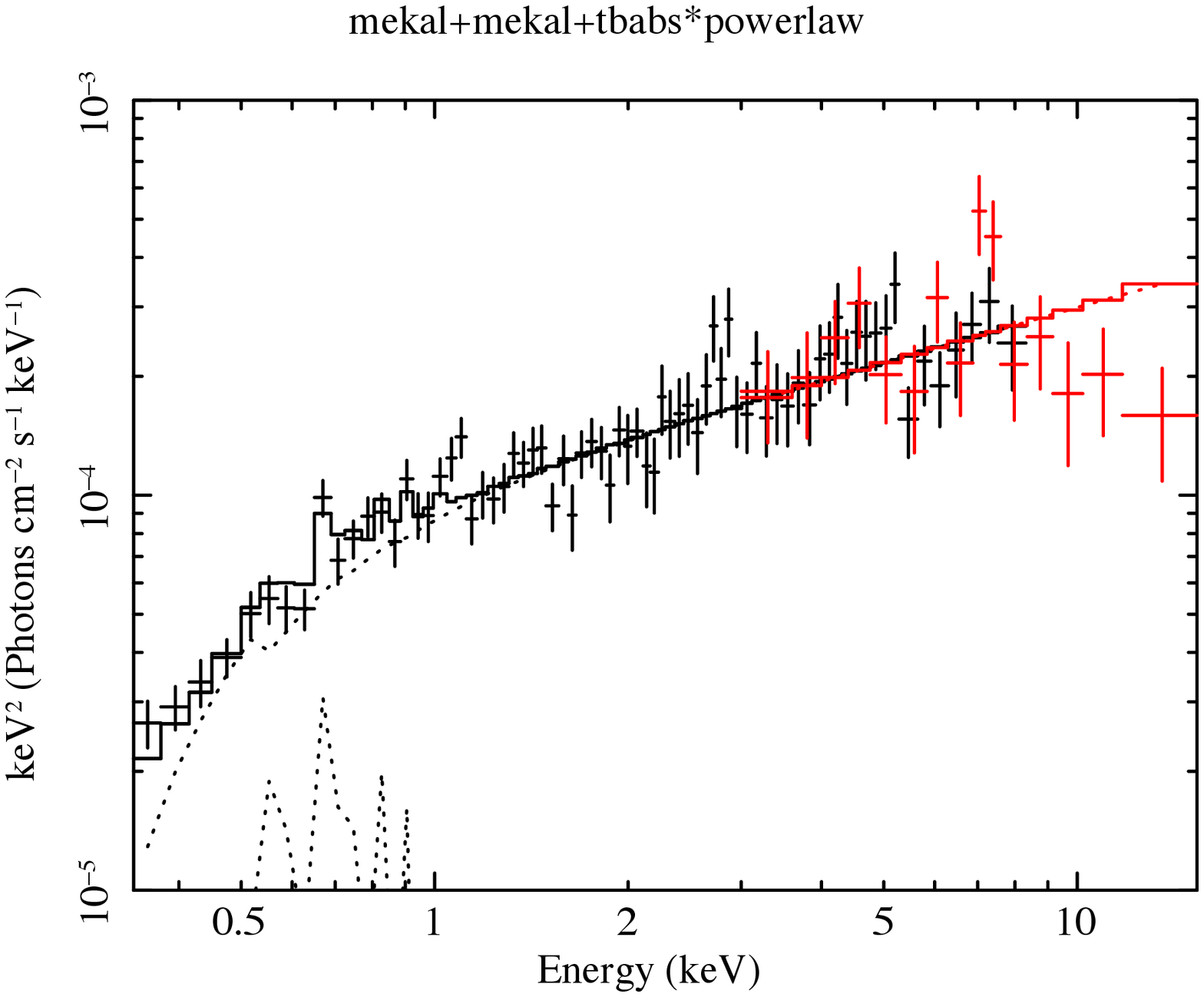}
\vspace{20px}
\includegraphics[width=80mm]{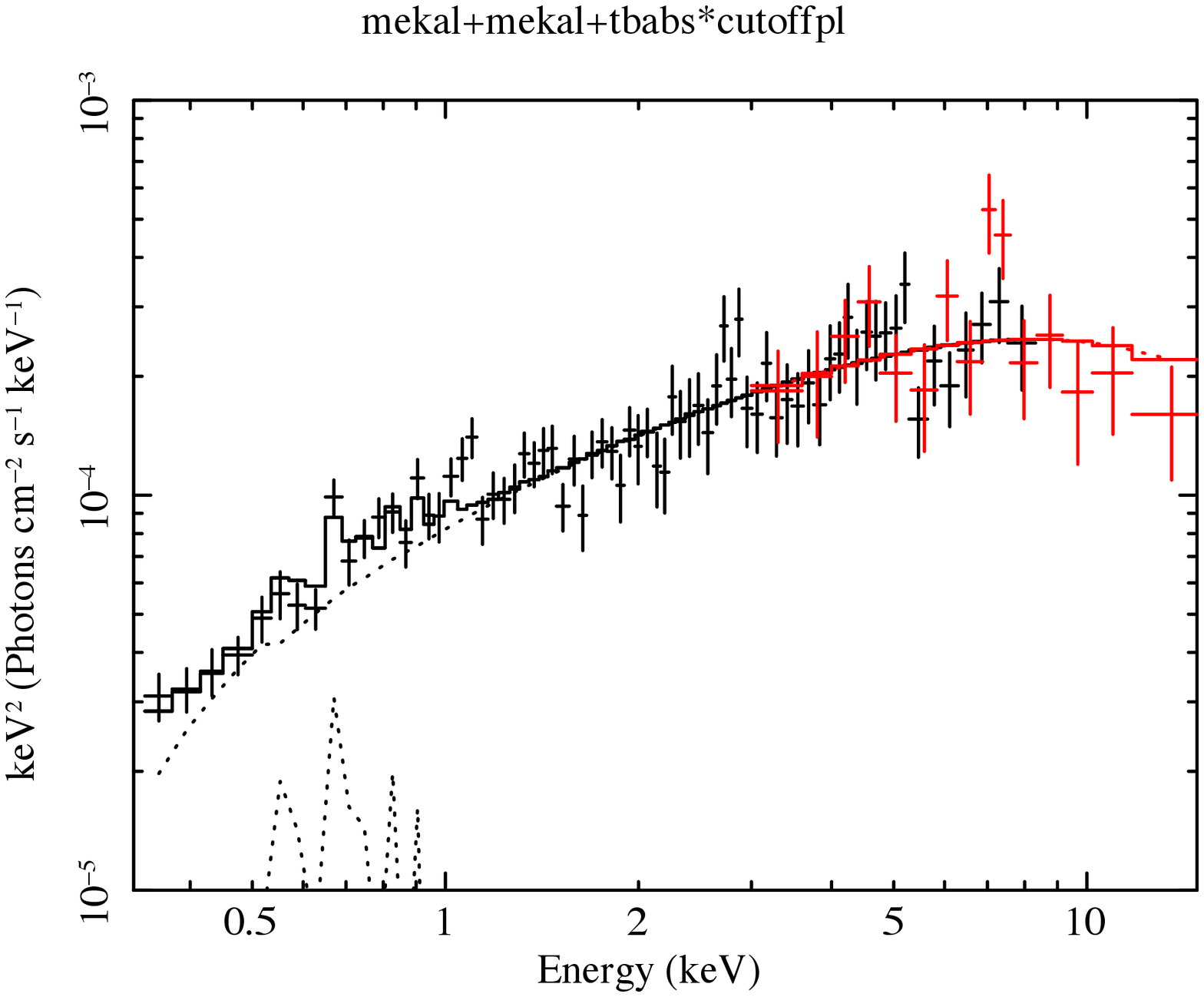}
\caption{The {\it NuSTAR} data (red) is fitted alongside the {\it XMM-Newton} data from observation 0303420201 (black; only EPIC-pn data is shown for clarity). The models used are {\tt mekal+mekal+tbabs*powerlaw} (top) and {\tt mekal+mekal+tbabs*cutoffpl} (bottom). The data are grouped into bins of at least 20 counts.}
\label{fig:xmmnu}
\end{figure}

ULX-7 is strongly detected by {\it NuSTAR} in the 3--24\,keV energy band. We extract a spectrum from a 40\,arcsec region around the source and fit it alongside the {\it XMM-Newton} observation nearest in flux (0303420201) with an absorbed power-law model as before, and also a cut-off power-law model ({\tt mekal+mekal+tbabs*cutoffpl}). An example of the joint {\it XMM-Newton} and {\it NuSTAR} fit for both models is shown in Fig.~\ref{fig:xmmnu}.

We find that in this case, the cut-off power-law model offers significant improvement over the simple power-law model, however this result comes with several caveats. First, while the data is flux-matched as well as possible the observations are not simultaneous, so we cannot be certain that the source has the same spectrum across both observations. There are also a number of fainter X-ray sources within the 40\,arcsec extraction region that may contaminate the hard emission. Finally, when fitted alongside the rest of the {\it XMM-Newton} observations, with a multiplicative constant to account for the difference in flux, the cut-off offers no significant improvement over the power-law.

We are therefore unable to definitively classify the accretion state of this source from the spectrum alone using current data, although we can say that it is almost certainly in either the sub-Eddington low/hard state or the hard ultraluminous state (with insufficient data quality to pick out the soft excess and turnover features to any great degree). Simultaneous deep observations of the source with {\it XMM-Newton} and {\it NuSTAR} will be required in order to better constrain the source spectrum.

\subsection{X-ray timing}

Over the twelve years that ULX-7 has been observed, its flux has been found to vary by well over an order of magnitude. Despite its very high dynamic range, it demonstrates little to no change in spectral state. In a similar way, all observations have very high amounts of variability, with 30--40\% rms. Additionally, its power spectra have consistent slopes between observations, so we combined the data from all observations for each telescope to improve statistics.

\begin{figure}
\centering
\includegraphics[width=80mm]{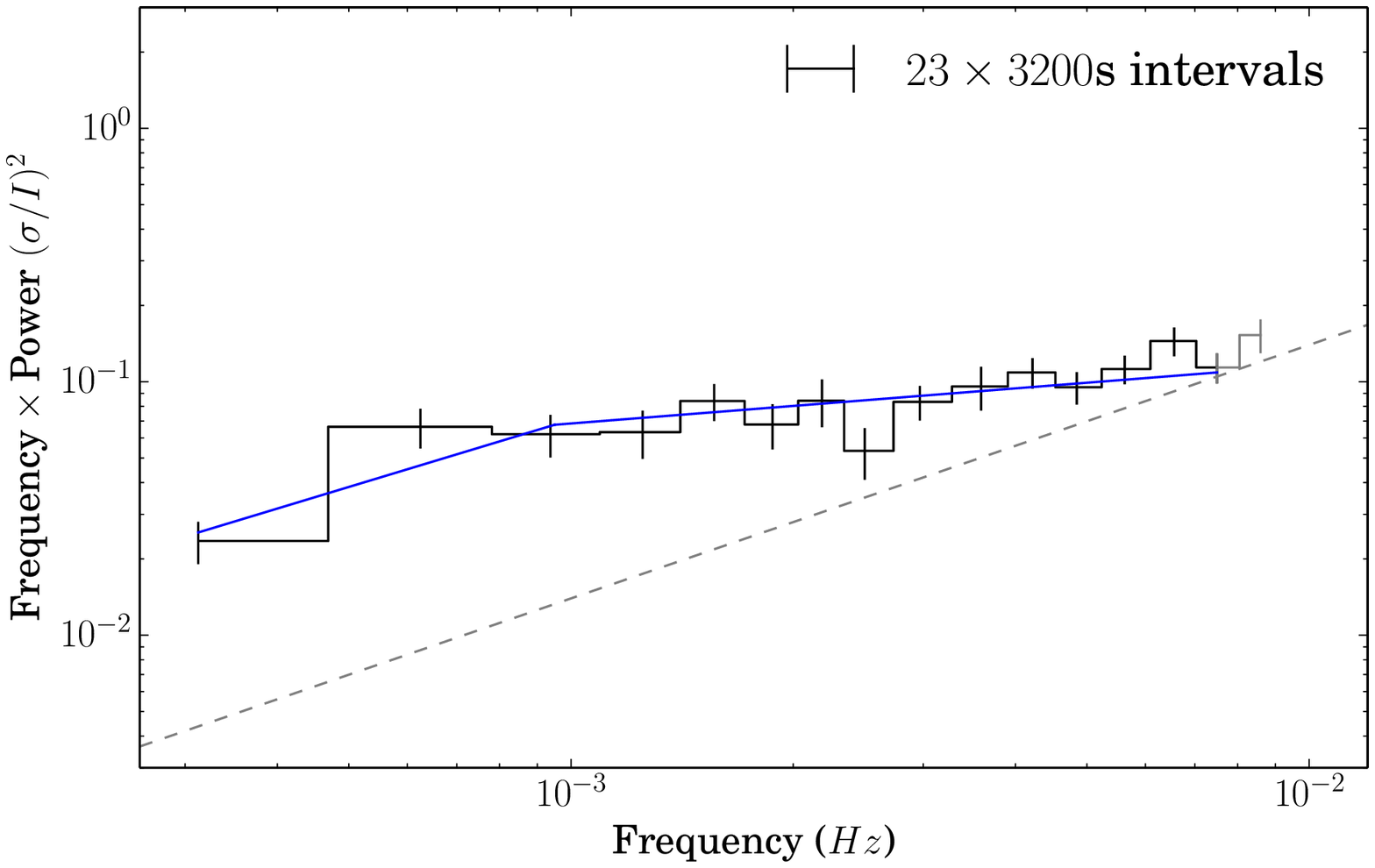}
\vspace{20px}
\includegraphics[width=80mm]{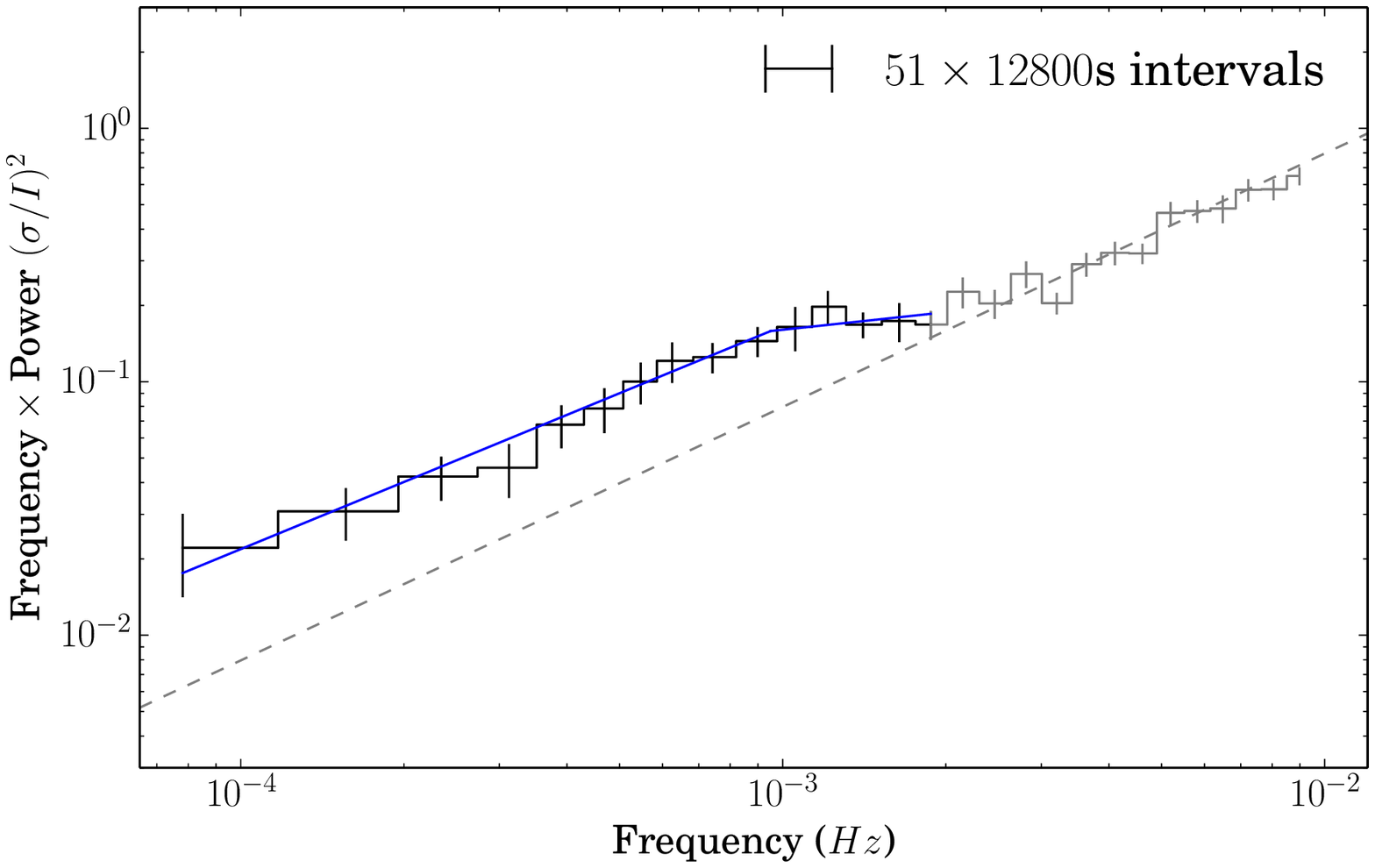}
\caption{The averaged power spectra for {\it XMM-Newton} (top) and {\it Chandra} (bottom). The white noise level in each case is given by the grey dashed line. The best-fit broken power-law model, detailed in the text, is shown in blue.}
\label{fig:powspec}
\end{figure}

Power spectra were constructed by dividing the light curves into equal length segments (3200\,s for {\it XMM-Newton} and 12800\,s for the longer {\it Chandra} observations) and averaging over the periodograms. While the {\it Chandra} data is noisier than that of {\it XMM-Newton} at short timescales, it does not have the same visibility restrictions, so observations could be taken over a much longer period of time. The two datasets are therefore very complementary and allow us to probe two orders of magnitude of frequency space. The resultant {\it XMM-Newton} and {\it Chandra} power spectra are shown in Fig.~\ref{fig:powspec}. Furthermore, this variability is consistently high across the entire 0.3--10\,keV energy range.

Attempting to fit all data above the white noise level with a single power-law results in a very poor fit. Therefore, we fit the {\it XMM-Newton} and {\it Chandra} data simultaneously using a broken power-law model. We find that this model is an excellent fit, with a break at $10^{-3}$\,Hz and best-fit slopes of $\alpha_1=0.1^{+0.5}_{-0.3}$ and $\alpha_2=0.8^{+0.6}_{-1.1}$, where $P(\nu)\propto\nu^{-\alpha}$. The slopes are not well-constrained, but a power spectrum break from $\alpha\sim0$ to $\alpha\sim1$ makes this shape analogous to the low frequency break found in the power spectra of BHs in the low/hard state, which can be approximated by a doubly-broken power-law. While this state cannot be directly confirmed as there is no evidence for a corresponding high frequency break, we can place a lower limit of $\sim8\times10^{-2}$\,Hz on a high frequency break from the white noise level of the {\it XMM-Newton} data.

We also perform an $H$-test \citep{dejager89} on the {\it XMM-Newton} EPIC-pn data in order to search for coherent pulsations. This would provide evidence towards the source being a pulsar, but we do not find any significant evidence of pulsations.

\section{Discussion}

Although ULX-7 is in the luminosity regime to be classed as a ULX, its behaviour is unusual compared to other sources of similar luminosity. While it has a hard spectrum, which may in itself be consistent with the hard ultraluminous state, it also exhibits high amounts of variability across the energy spectrum. High variability is only expected in the soft ultraluminous state, and then only in the hard emission component. Therefore, since its behaviour is not consistent with the majority of ULXs found in ultraluminous accretion states, we examine a number of other scenarios for explanations of its properties.

\subsection{Background AGN}

Occasionally, IMBH candidates will turn out to be background AGNs coincident with the foreground galaxy. The optical results can be used to verify the source's location as being within the host galaxy. Firstly, the colours and magnitudes of the potential counterparts are all consistent with being OB supergiants, suitable companion stars for a HMXB, or else clusters of lower mass stars. Also, the X-ray/optical flux ratios are higher than would be expected for a background AGN. Therefore we can be reasonably confident that ULX-7 is not a background source but genuinely located within its host galaxy.

Furthermore, we observe strong variability at frequencies higher than expected for a background AGN, which is another point against this interpretation.

\subsection{Neutron Star}

With the discovery that M82 X-2 is in fact a neutron star ULX \citep{bachetti14}, one possible reason for ULX-7's unusual properties is that it may be a neutron star rather than a BH binary. The key evidence for this scenario would be the detection of coherent pulsations, which would point to the source being a pulsar, however we do not detect any in the {\it XMM-Newton} data.

Comparing it with another well-known neutron star source, the very luminous Z-source LMC X-2, we find that ULX-7's spectrum is softer. Furthermore, the power spectrum does not exhibit the steep slope ($\alpha=1-2$) at very low frequencies found in Galactic Z-sources \citep{hasinger89}. Therefore, while we cannot rule out a neutron star interpretation, this source appears to be inconsistent with what we would expect from a high-luminosity Z-source.

\subsection{Intermediate-mass Black Hole}

An alternate explanation is that ULX-7 is a BH accreting in a low/hard state. This scenario is supported by the consistently hard source spectrum with $\Gamma\sim1.5$, high amounts of variability across all energies, and a power spectrum exhibiting a break analogous to the low frequency break found in lower luminosity BH binaries in the hard state. Its high luminosity would therefore imply a high BH mass, making it a candidate IMBH.

If we assume that this is the case, we are able to place upper limits on the BH mass using the radio and timing data. We would expect to see radio emission from the jets of a source in the hard state, but as we make no radio detection, we can derive an upper limit on the 1.5\,GHz flux density of 87\,$\mu$Jy\,beam$^{-1}$. We are then able to use the fundamental plane in BH mass, X-ray luminosity and radio luminsity to place a limit on the mass of this object. We use the fundamental plane equation with the least scatter ($\sigma=0.12$) in \citet{kording06}, $\log L_{\rm X} = 1.59\log L_{\rm 5\,GHz} - 1.02\log M_{\rm BH} - 10.15$, and assume a flat radio spectral index of $\alpha=0.15$ to find $L_{\rm 5\,GHz}$. In this way, we find an upper limit on the BH mass of $M_{\rm BH}<1.95\times10^5$\,M$_{\odot}$.

We can find another, independent upper limit by using the relationship between the high frequency break in the power spectrum and the BH mass, found to be a relation that applies to AGNs and BH binaries alike \citep{mchardy06}. While we are unable to observe a high frequency break, we can use the lower frequency limit of such a break, $8\times10^{-2}$\,Hz, to place an upper limit on the mass using the equation $T_B = 2.1\log M_{\rm BH} - 0.98\log L_{\rm bol} - 2.32$. We find an upper limit of $M_{\rm BH}<9.12\times10^4$\,M$_{\odot}$.

Both of these upper limits on the BH mass are consistent with the source being an IMBH.

The points against this interpretation are the results of the {\it NuSTAR} observation that suggest a turnover at high energies, as well as hints of a soft excess seen in the longest {\it XMM-Newton} observation, 0303420101. These features would be evidence of a source in the hard ultraluminous state, albeit with unusual variability properties, or even a yet more uncommon accretion state that is yet to be characterised. 

However, the low significance of the turnover as a feature, coupled with the fact that the {\it NuSTAR} observation is not contemporaneous with any of the other X-ray observations means that we are unable to say conclusively whether the source is in a two-component ultraluminous state or not. A simultaneous deep observation of M51 using both {\it XMM-Newton} and {\it NuSTAR} would allow the spectrum to be better constrained and to confirm or rule out an IMBH interpretation.

\section{Summary}

The construction of a catalogue of ULXs using 3XMM-DR4 data has led to the discovery that M51 ULX-7 has unusual spectral and variability properties for a ULX, with a consistently hard spectrum and high amounts of variability that do not easily place it within an ultraluminous state classification. 

Optical data from {\it HST} confirm that it is very likely to genuinely reside within M51 as opposed to being a background AGN. The lack of coherent pulsations coming from the source, as well as dissimilarities with Galactic neutron stars, suggest that it is probably a BH as opposed to a neutron star ULX like M82 X-2. 

Examination of the {\it XMM-Newton} and {\it Chandra} spectra show the source to have a consistently hard spectrum ($\Gamma\sim1.5$) after accounting for contamination by soft diffuse emission, despite large variation in flux. The power spectrum also appears to feature a break analogous to the low frequency break found in the low/hard state in BH binaries. These features would suggest an IMBH accreting in the hard state, and upper limits on the BH mass found using the lack of a radio detection and high frequency break in the power spectrum are consistent with this interpretation.

However, the {\it NuSTAR} data shows tentative evidence of a spectral turnover at high energies, which would be evidence towards an ultraluminous state classification, although simultaneous observations with both {\it XMM-Newton} and {\it NuSTAR} will be required to confirm the existence of this feature and come to a more decisive conclusion on the nature of ULX-7. 

\acknowledgements
We gratefully acknowledge support from the Science and Technology Facilities Council (HE through grant ST/K501979/1).

\end{document}